\documentclass[reprint, amsmath, amssymb, aps, superscriptaddress, prb, longbibliography]{revtex4-2}

\usepackage{graphicx}% Include figure files
\usepackage{epstopdf}
\usepackage{dcolumn}% Align table columns on decimal point
\usepackage{bm}% bold math
\usepackage{color}
\usepackage[colorlinks=true, urlcolor=blue, linkcolor=blue, citecolor=blue]{hyperref}
\usepackage{natbib}
\usepackage{amsbsy}
\usepackage{lipsum}
\usepackage{verbatim} 

\usepackage[dvipsnames]{xcolor}
\usepackage{pst-all}
\usepackage{ulem}

\usepackage{lineno}

\begin{document}

\title{Quantum Hall effect and zero plateau in bulk HgTe}

\author{M.\,L.\,Savchenko}%\email{mlsavchenko@isp.nsc.ru}
\affiliation{Institute of Solid State Physics, Vienna University of
	Technology, 1040 Vienna, Austria}

 \author{D.\,A.\,Kozlov}
\affiliation{Experimental and Applied Physics, University of Regensburg, D-93040 Regensburg, Germany}
\affiliation{Institute of Semiconductor Physics, 630090 Novosibirsk, Russia}

\author{S.\,S.\,Krishtopenko}
\affiliation{Laboratoire Charles Coulomb (L2C), UMR 5221 CNRS-Université de Montpellier, Montpellier, France}

\author{N.\,N.\,Mikhailov}
\affiliation{Institute of Semiconductor Physics, 630090 Novosibirsk, Russia}
%\affiliation{Novosibirsk State University, Novosibirsk 630090, Russia}

%\author{S.\,A.\,Dvoretsky}
%\affiliation{Institute of Semiconductor Physics, 630090 Novosibirsk, Russia}

\author{Z.\,D.\,Kvon}
\affiliation{Institute of Semiconductor Physics, 630090 Novosibirsk, Russia}
\affiliation{Novosibirsk State University, Novosibirsk 630090, Russia}

\author{A.\,Pimenov}
\affiliation{Institute of Solid State Physics, Vienna University of
	Technology, 1040 Vienna, Austria}

\author{D.\,Weiss}
\affiliation{Experimental and Applied Physics, University of Regensburg, D-93040 Regensburg, Germany}

\date{\today}

\begin{abstract}
The quantum Hall effect, which exhibits a number of unusual properties, is studied in a gated 1000-nm-thick HgTe film, nominally a three-dimensional system. 
A weak zero plateau of Hall resistance, accompanied by a relatively small value of $R_{xx}$ of the order of $h/e^2$, is found around the point of charge neutrality. 
It is shown that the zero plateau is formed by the counter-propagating chiral electron-hole edge channels, the scattering between which is suppressed. 
So, phenomenologically, the quantum spin Hall effect is reproduced, but with preserved ballisticity on macroscopic scales (larger than 1\,mm). 
It is shown that the formation of the QHE occurs in a two-dimensional (2D) accumulation layer near the gate, while the bulk carriers play the role of an electron reservoir. 
Due to the exchange of carriers between the reservoir and the 2D layer, an anomalous scaling of the QHE is observed not with respect to the CNP, but with respect to the first electron plateau.
\end{abstract}

\maketitle

\section{Introduction}

The quantum Hall effect (QHE)~\cite{Klitzing1980} is one of the brightest hallmarks of physics in second half of 20-th century. 
Just QHE led to the birth of conductance quantum ($e^2/h$) reflecting quantum mechanics laws in clearest, simplest and most accurate way. 
So, the deep and width of QHE physics are inexhaustible and the attention to its study from both theoretical and experimental points of view is saved for all time. 
The main questions, concerning QHE at the present time, are connected to its realization in such novel systems as graphene~\cite{Checkelsky2008}, semimetals~\cite{Mendez1985, Gusev2012a}, three-dimensional topological insulators~\cite{Yoshimi2015, Xu2014a, Ziegler2020}. 
In graphene and semimetals, zero Landau level (LL) manifesting by zero Hall conductivity $\sigma_\text{xy} = 0$ is of much interest. 
There are several theoretical approaches for the issue~\cite{Abanin2007, DasSarma2009, Khveshchenko2001}, but there is still no a general approach for the problem of realization zero plateau. 
Moreover, to the date we do not know any works, where this plateau would be observed in measured entity $\rho_\text{xy}$, not in recalculated one, such as $\sigma_\text{xy}$. 

Three-dimensional topological insulators are systems, where topologically non-trivial and non-degenerate surface states exist at all Fermi level positions~\cite{Ando2013, Bansil2016}. 
If the bulk of such materials is insulating, they should realize QHE, because conduction surfaces of such structure are natural two-dimension systems. 
There are works in this area, where authors claim that they observe QHE, originating from these surface states~\cite{Xu2014a, Yoshimi2015, Brune2011}. 
This question still needs to be checked, and another curious question here is the possibility to observe half-integer QHE~\cite{Konig2014} that was not unambiguously presented to the date. 
Moreover, it was claimed~\cite{Wang2017a} that QHE can be realized in three-dimension topological semimetals. 
And recently it has been demonstrated in bulk ZrTe$_5$~\cite{Tang2019}, in fact having the layer structure. 
Since it is completely new view on the QHE formation, more experiments in the feld are highly expected, especially if they can be realized on the base of well-known structures.

Here we present an experimental study of a high-quality gated 1000-nm-thick HgTe film, which is a zero-gap semiconductor with an inverted band structure and thus hosts both bulk carriers and topological surface states. In our previous work \cite{Savchenko2023} we studied classical magnetotransport and Shubnikov-de Haas (SdH) oscillations of this structure. It was shown that  there are low-mobility 3D carriers and high-mobility 2D carriers, whose charge depends on the gate voltage. The 2D carriers are located at the interface between HgTe and CdHgTe (topological electrons) or in its vicinity (trivial electrons or holes, or Volkov-Pankratov electrons), in the accumulation layer formed by the gate voltage. Both 2D electrons and holes exhibit pronounced SdH oscillations, which are sensitive to the perpendicular component of the magnetic field. The 3D electrons are located in the bulk and act as a separate classical conductivity channel.  In the current work, we  focus on quantizing magnetic fields and show a distinct QHE with several peculiarities, including the formation of the $\rho_{\rm xy}$ zero plateau, quasi-ballistic edge channels on a macroscopic scale, and anomalous scaling with respect to the first electron plateau. The observed features are explained by the coexistence of 2D and 3D carriers, where the former are responsible for the formation of the QHE and the latter play a role of charge reservoir.

\section{Methods}
\begin{figure}
	\includegraphics[width=0.7\columnwidth]{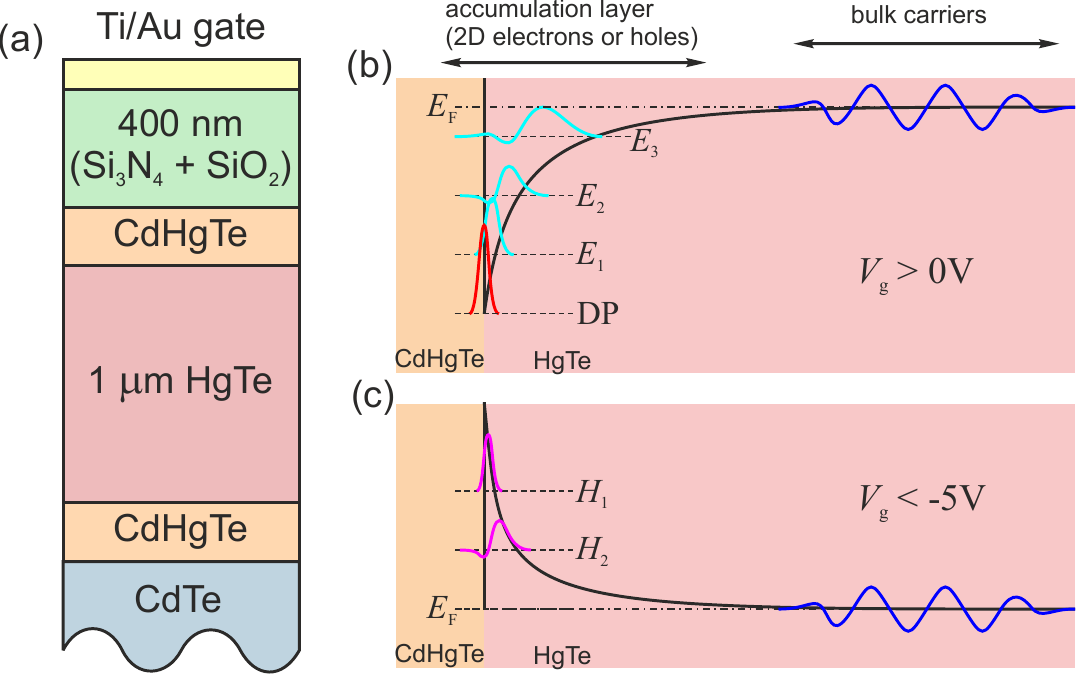}
	\caption{
        (a) Schematic cross-section of the structure under study. 
        (b) and (c) 
        The suggested band diagram of the accumulation layer for positive and negative gate voltages. 
        $E_i$ and $H_i$ denotes edges of electron and hole sub-bands in the accumulation layer, DP is the Dirac point of topological surface electrons, $E_F$ is the Fermi level.
	} \label{Figbands}
\end{figure}

Measurements are carried out on 1000\,nm~HgTe films that have been grown by molecular beam epitaxy on a GaAs(013) substrate at the same conditions and with the same layer ordering as it was for usual 80 and 200\,nm films~\cite{Kozlov2014, Savchenko2019}.
In Fig.~\ref{Figbands}\,(a) we schematically show cross-section view of the system under study.
The 1\,$\mu$m~HgTe film is placed between thin Cd$_{0.6}$Hg$_{0.4}$Te buffer layers, a Ti/Au gate has been deposited on the 200+200\,nm of Si$_3$N$_4+$SiO$_2$ insulator grown by a low temperature chemical vapor deposition process.
An approximate thickness of the pseudomorphic growth of a HgTe film on a CdTe substrate with a 0.3\%
larger lattice constant is about 100 -- 150\,nm  (according to~\cite{Brune2011} and our experience).
Thereby we believe that our ten times thicker HgTe film is fully relaxed to its own lattice constant and is a zero gap semiconductor 
%has some meV of the bands overlap 
\cite{Berchenko1976}.

The studied Hall-bars have a 50\,$\mu$m current channel and equal to 100 and 250\,$\mu$m distances between potential probes. 
Transport measurements were performed using a standard lock-in technique with a driving current in the range of 10$^{-10}$ -- 10$^{-7}$\,A in a perpendicular magnetic field $B$ up to 12\,T in the temperature range of 70\,mK -- 10\,K. 
A typical frequency for transport measurements was 12\,Hz that was decreased to 2\,Hz in the case of high-magnetic fields.

\section{Results}

\begin{figure}
	\includegraphics[width=1\columnwidth]{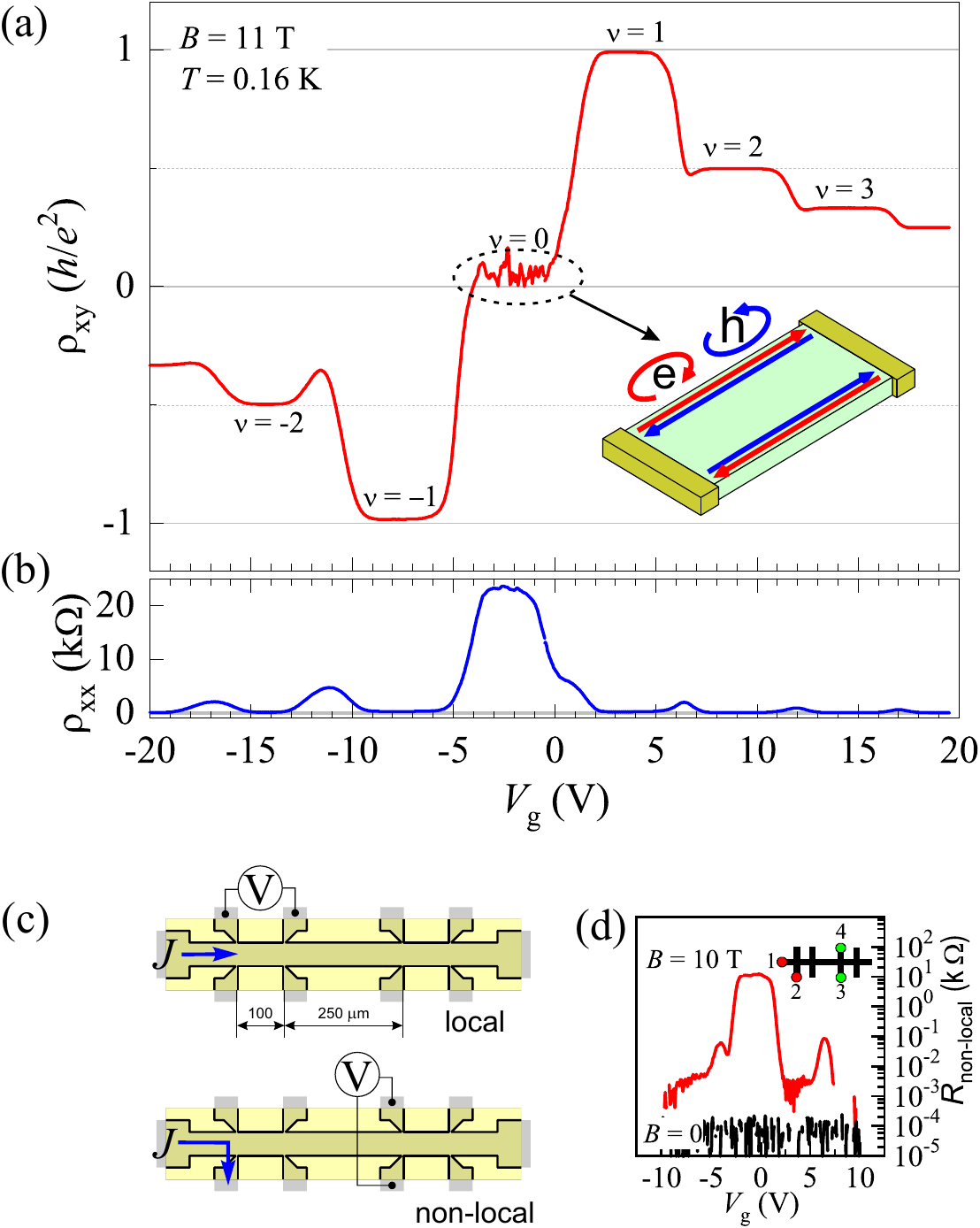}
	\caption{
        (a) and (b) QHE plateaus in Hall resistance $\rho_\text{xy}(V_g)$ and minima in longitudinal resistance  $\rho_\text{xx}(V_g)$ measured at $B = 11$\,T and temperature $T = 0.16\,$K. 
		Note the formation of zero plateau in $\rho_\text{xy}$. 
        Inset:	the schematic representation of counter-propagating chiral electron and hole edge channels.
		(c)~The macroscopic hallbar sketch with two schemes of measurement: local (top, used for $\rho_\text{xy}$ and $\rho_\text{xx}$) and non-local (bottom). 
        (d)~Non-local resistance versus gate voltage measured at $B=0$ (black) and $B=10$\,T (red) at $T=0.16$\,K. 
        Inset: the measurement scheme, red and green dots represent the current and voltage probes, respectively.
	} \label{Fig1}
\end{figure}

In~Fig.~\ref{Fig1}\,(a) we show the main experimental result of our work:  the Hall resistance $\rho_\text{xy}$ versus gate voltage $V_\text{g}$ measured at $B = 11$\,T demonstrates clear plateaus both on the electron (at positive $V_\text{g}$) and the hole side (at negative $V_\text{g}<-5\,$V). The quantized values of $\rho_\text{xy}$ are equal to $h/e^2$ and its integer fractions. The plateaus in $\rho_\text{xy}$ are accompanied by close to zero values in the diagonal resistivity $\rho_\text{xx}(V_\text{g})$ in Fig.~\ref{Fig1}\,(b). 
Taken together, this clearly indicates the formation of a pronounced QHE, which  observation in a 3D system may be unexpected.
However, the carriers that contribute the most to the conductivity are actually 2D carriers 
constituting from topological surface states and trivial carriers
in the accumulation layer near the gate~\cite{Savchenko2023}, see~Fig.~\ref{Figbands}~(b) and (c). 
The dominant role of two-dimensional carriers in the formation of the QHE is also evidenced by the sensitivity of the filling factor to the gate voltage, as well as the equidistance of the plateaus.
On the other hand, it is unexpected that the large number of bulk carriers does not prevent the QHE formation or somehow distort it.

At first glance, the behavior of $\rho_\text{xx}$ and $\rho_\text{xy}$ in Fig.~\ref{Fig1}~(c) and Fig.~\ref{Fig1}~(d) is similar to that observed in other electron-hole systems such as graphene~\cite{Abanin2007, Checkelsky2008, Zhang2006, Hong2021} or in closely related HgTe quantum wells~\cite{Kvon2011a, Gusev2012a, Gusev2017, Ziegler2020, Buttner2011, Jost2017, Kozlov2023DF}. 
However, there is a fundamental difference between these and our results: 
near the charge neutrality point (CNP), we observe a zero plateau $\rho_\text{xy}$  accompanied by a relatively small value of $\rho_\text{xx} \approx h/e^2$. 
In the quantum Hall regime, the minima of $\rho_\text{xx}$ usually coincide with the minima of $\sigma_\text{xx}$. At the CNP, however, this reciprocity is violated: as the magnetic field increases and the temperature decreases, the value of $\rho_{\rm xx}$ typically becomes many times larger than $h/e^2$, while $\rho_{\rm xy}$ remains of the order of $h/e^2$~\cite{Abanin2007,Jiang2007,Zhang2006,Checkelsky2008,Raichev2012b,Gusev2012a,Kozlov2014b, Kozlov2014a,Gusev2012,Gusev2017, Lygo2023, Guo2022, Nichele2014}. Hence $\rho_{\rm xx} \gg \rho_{\rm xy}$, both $\sigma_\text{xx}$ and $\sigma_\text{xy}$ naturally tend to zero during the inversion of the resistivity tensor. 
Although the zero  $\sigma_\text{xy}$ plateau is also observed in the system under study (see 
%Sec.~\ref{fig_SM_QHE_and_Non-local_Sigma} 
Fig.~S3
of the Supplemental Material (SM)~\cite{SM_universal}), we emphasize the unusualness and importance of the first, to our knowledge, experimental observation of a zero $\rho_\text{xy}$ plateau. 
The observed  
plateau, unlike plateaus with other filling factors, is "weak": first, the plateau appears only after the procedure of antisymmetrization of the experimental data (see  
Fig.~S4
in SM~\cite{SM_universal}). The need for antisymmetrization is not surprising, since the measured value of $\rho_\text{xy}$ is always complemented by a field-symmetric signal proportional to $\rho_\text{xx}$, which reaches its maximum in the CNP. 
Second the zero plateau does not have a precise  quantization accuracy
since the temperature is not exactly the same at positive and negative $B$, modifying the $\rho_{xx}$ impact, and hence, measured $\rho_{xy}$ values.
It results in non-zero values of the Hall resistance and random but reproducible fluctuations around $\nu = 0$, which amplitude scales with temperature variations,
see Fig.~\ref{Fig1}~(a), \ref{Fig3}~(d), and 
%Sec.~\ref{sec: QHE_diff_T} 
Sec.~S5
of SM~\cite{SM_universal}. 
Despite these features, which indicate the special nature of the zero plateau, its width and slope are consistent with the other plateaus.

The value of resistivity $\rho_\text{xx} \lesssim h/e^2$ at the CNP, as well as its weak temperature dependence at $T<0.2$\,K (see 
%Sec.~\ref{sec: QHE_diff_T} 
Fig.~S5
in SM~\cite{SM_universal}),
indicates the character of conductivity corresponding to a disordered metal. However, this conductivity can be due to both bulk and edge mechanism. In the first case the conductivity is formed as a sum of partial conductivities (electron and hole contributions are added for $\sigma_{\rm xx}$ component, but subtracted for $\sigma_{\rm xy}$ due to opposite charges), and in order to obtain a zero plateau $\rho_{\rm xy}$ there should be a full compensation of the Hall conductivity $\sigma_{\rm xy}$. In the second case the counter-propagating electron and hole channels are formed along the edges of the sample. 
To determine which of these models is most relevant to the system under study,
%To answer the question of which of these models is relevant to the system under study, 
we probed its non-local transport response. 
Such an approach has been widely used in order to prove the edge transport of both 2D TIs~\cite{Roth2009, Gusev20112DTI, Brune2012, Gusev2013InPlane, Rahim2015, Olshanetsky2015, Piatrusha2019, Ryzhkov2022} and non-topological systems in the QHE regime~\cite{Nichele2014, Morimoto2015, Gusev2017, Nachawaty2018, Brown2018, Hong2021}. 
%To do this, we 
%passed current across the sample and 
During non-local measurements, we
separated the current (contacts 1 and 2 in Fig.~\ref{Fig1}~(b) and (c)) and voltage (3 and 4) pairs of the contacts by 350\,$\mu$m.
The results of $R_{1-2,3-4}(V_g)$ measurements are shown for magnetic field of 0 and 10\,T in Fig.~\ref{Fig1}~(c)).
The peak of non-local resistance in the CNP at 10 Tesla (red curve) is clear evidence for the edge-based nature of the transport. On the contrary, the dominance of bulk conduction at zero magnetic field leads to an exponentially small value of the measured signal (black curve). 
The experiment clearly shows that near the zero plateau, %edge transport dominates and the contribution of 
the bulk conductivity is negligible, 
while the counter-propagating electron and hole channels dominate the transport response.

Although edge transport is not unique for the 2D electron-hole systems in the vicinity of the CNP and in the QHE regime \cite{Gusev2012a, Abanin2007, Hong2021, Gusev2017}, the observed low value of the diagonal resistance is rather unusual. Typically, electron-hole edge channels in both QHE systems and 2D TIs show a lack of the backscattering protection on a scale larger than a few $\mu$m, leading to a value of the resistance much higher than $h/e^2$. In this context, the  observed value of $\rho_{\rm xx} \approx h/e^2$ (the raw resistance $R_{\rm xx}$ is 2 times larger) indicates a quasi-ballistic character of the transport, preserved for unknown reasons on a scale of hundreds of microns. An independent confirmation of the quasi-ballistic character is also provided by the value of the maximum of the non-local resistance ($\approx10$\,kOhm), which coincides with the theoretical estimate of $(2/5)h/e^2$ for ballistic edge channels in the used measurement geometry \cite{Olshanetsky2015}. On the other hand, the absence of a zero plateau $\rho_{\rm xy}$ before antisymmetrization in the magnetic field and its high sensitivity to temperature indicate that protection from backscattering is also not complete. More research is needed in this area.

\begin{figure}
	\includegraphics[width=1\columnwidth]{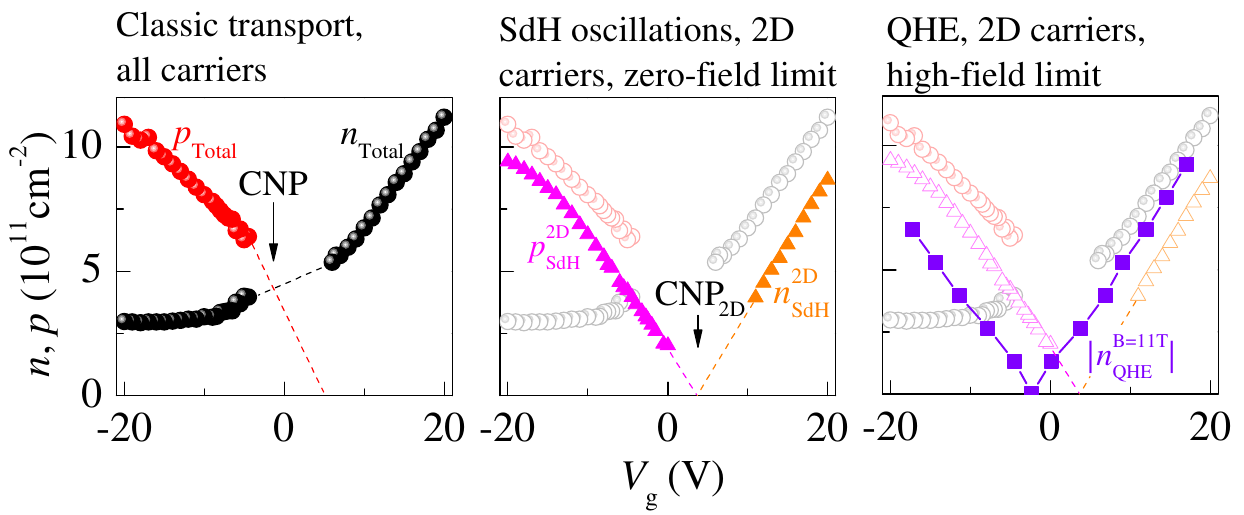}
	\caption{
		Comparison of the charge densities obtained by different methods. 
		(a)~The gate voltage dependences of total electron $n_\text{Total}$ and hole $p_\text{Total}$ densities obtained from low-field classic magnetotransport.
		(b)~2D electron $n_\text{SdH}^\text{2D}$ and ``hole'' $(p-n)_\text{SdH}^\text{2D}$ densities obtained from the analysis of SdH oscillations (filled symbols).
		(c)~QHE density $n_\text{QHE}^\text{B=11T}$ obtained from maxima and minima positions of $\rho_\text{xx}$ (filled symbols).
		The difference between total $n_\text{Total}$ and 2D electron $n_\text{SdH}^\text{2D}$ densities comes from the existence of conducting 3D bulk electrons. 
		While the QHE density $n_\text{QHE}^\text{B=11T}$ lies between 2D and 3D zero-field electron densities indicating that some 3D electrons acquire a 2D nature at high fields.
	} \label{Fig2}
\end{figure}

We investigate the role of bulk 3D carriers in a formation of the QHE state. For this purpose, we compare the carrier density obtained by different methods. Fig.~\ref{Fig2}~(a) shows the total density (i.e. both 2D and 3D carriers) of electrons $n_{\rm Total}$ and holes $p_{\rm Total}$ obtained from classical magnetotransport~\cite{Savchenko2023}. 
Panel~(b) shows the electron $n_{\rm SdH}^{\rm 2D}$ and differential $(p-n)_{\rm SdH}^{\rm 2D}$ density of 2D carriers only, obtained from the SdH oscillations analysis~\cite{Savchenko2023}. 
Finally, panel (c) shows the density of the carriers extracted from the centers of the $\rho_{\rm xy}$ plateaus in Fig.~\ref{Fig1}~(c) using the formula $n_{\rm QHE} = |\nu B e/h|$, where $\nu$ is the filling factor, $B e/h$ is a Landau level degeneracy, and $B = 11$\,T. 
%The resulting trace has the same slope as $n_{\rm SdH}^{\rm 2D}$ and $(p-n)_{\rm SdH}^{\rm 2D}$ with a slight offset between them. 
The the gate voltage dependences of $n_{\rm SdH}^{\rm 2D}$ and $(p-n)_{\rm SdH}^{\rm 2D}$ have nearly the same slope
confirming an assumption that the filling factor is determined by the total charge density of 2D carriers. 
This is possible with a complete hybridization of various 2D carriers (i.e. topological electrons with other carriers in the accumulation layer), as observed in thinner films of HgTe~\cite{Ziegler2020, Kozlov2016, Savchenko2019} as well as in other systems~\cite{Karalic2019}. 
The reason for the gate voltage shift between $n_{\rm SdH}^{\rm 2D}$ and $(p-n)_{\rm SdH}^{\rm 2D}$ is explained further.
%\textcolor{red}{so maybe move this paragraph down?}

%
\begin{figure}
	\includegraphics[width=1\columnwidth]{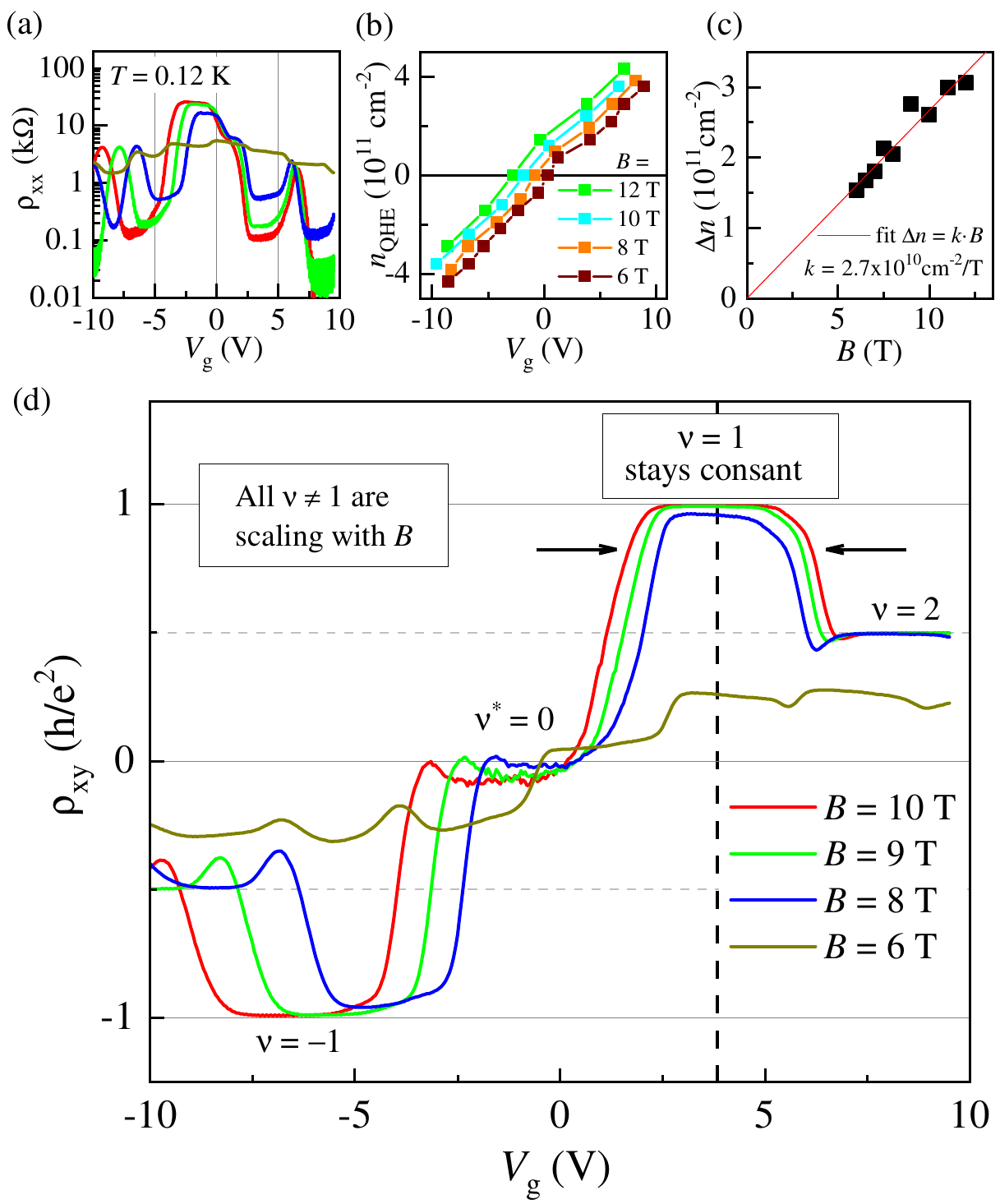}
	\caption{
        (a) and (d)~The gate voltage dependences of longitudinal and Hall resistance in the QHE regime at different magnetic fields.
        The position of the first $\rho_\text{xx}$ minimum and the corresponding $\rho_\text{xy}$ plateau ($V_g = 4$\,V) remains the same regardless of the magnetic field, while all other minima and plateaus scale around $\nu=1$ as $B$ changes.
		(b)~The gate voltage dependence of the charge density $n_\text{QHE}$ obtained from the  maxima and minima positions of $\rho_\text{xx}$ for $B = 6, 8, 10$ and 12\,T. 
        Note the linear $n_\text{QHE}(V_g)$ dependence% of the density on the gate voltage
        , as well as the linear shift between the curves in the magnetic field.
		(c)~The dependence of the gate voltage averaged charge density shift $\Delta n(B) \equiv n_\text{QHE}(B) - n_{\rm SdH}^{\rm 2D}$ taken from the panel (b). Extrapolating to $B=0$ gives $\Delta n=0$, i.e. the coincidence of $n_\text{QHE}$ and $n_{\rm SdH}^{\rm 2D}$. %Red line: fit $\Delta n = k \cdot B$ with $k = 2.7\times10^{10}$\,cm$^{-2}$/T.
  	} \label{Fig3}
\end{figure}

We analyze the QHE evolution in the magnetic field. Fig.~\ref{Fig3}(a) and (d) show the $\rho_{\rm xx}(V_g)$ and $\rho_{\rm xy}(V_g)$ dependences, respectively, in a magnetic field from 6 to 10\,T. Starting from a magnetic field of 8\,T, distinct  $\rho_{\rm xy}$ plateaus with filling factors from -2 to +2 are observed, accompanied by deep and flat minima of $\rho_{\rm xx}$. 
For $\nu = 0$ the $\rho_{\rm xx}$ exhibits a broad flat top maximum with a value of 15-25\,kOhm. 
At lower magnetic fields, $B = 6$\,T, there is an abrupt collapse of QHE: instead of gradually blurring the plateaus, their position along the Y-axis changes, becoming chaotic, while the flat areas are replaced by inclined ones. At the same time, the $\rho_{\rm xx}(V_g)$ dependence shows a transition from oscillations typical of the QHE to an almost monotonic curve characteristic of classical transport. Both curves retain a strict periodicity along the $V_g$ axis, with a period that is 6/10 of the period at $B=10$\,T. Taken together, this indicates that the 2D carriers' spectrum is still quantized at 6\,T, but the measured dependences $\rho_{\rm xx}(V_g)$ and $\rho_{\rm xy}(V_g)$ turn out to be distorted by the contribution of the 3D carriers, whose classical conductivity is summed up with the conductivity of 2D carriers. The conductivity of the 3D carriers is independent of the gate voltage because of the screening, but appears to decrease rapidly with increasing magnetic field. In a magnetic field of 8\,T or more, the 3D carriers localize and no longer significantly distort the $\rho_{\rm xx}$ and $\rho_{\rm xy}$ traces. However, the influence of bulk carriers can still be detected at $B=8$\,T by the deviation of the position of the $\rho_{\rm xy}$ plateaus with respect to $h/\nu e^2$, as well as by the insufficiently deep but saturated minima of the $\rho_{xx}$. From the depth of the minima (and assuming that the partial conductivity of the 2D carriers in the gap between LLs is negligibly small) we can estimate that the bulk conductivity drops between $10^{-6}$ to $10^{-7}$\,Ohm$^{-1}$ with increase of magnetic field from 8 to 10\,T. 

Another striking anomaly of the study is related to the position of the $\rho_{\rm xy}$ plateaus and $\rho_{\rm xx}$ minima along the $V_g$-axis.
These points are associated with an integer number of filled LLs, 
when the density is determined by a standard QHE equation $n_{\rm QHE} = |\nu eB/h|$. Because of the linear relationship between density and gate voltage, the distance between adjacent plateaus is constant and proportional to the magnetic field, and normally the curves $\rho_{\rm xy}$ and $\rho_{\rm xx}$ scale with the center at the CNP \cite{Checkelsky2008, Zhang2006, Hong2021, Kvon2011a, Gusev2012a, Ziegler2020, Buttner2011}. However, the traces shown in Fig.~\ref{Fig3}~(a) and (d) behave differently: as the magnetic field is varied from 6 to 10\,T, the position of the first electron plateau located at $V_g = 4$\,V remains unchanged, while the position of the other plateaus (including the zero one) scales linearly with the magnetic field relative to the first one. Thus, for some reason, the first plateau plays the role of the gate voltage coordinate center. On the other hand, the electron density at the center of the first plateau is equal to $e B /h$, i.e. the electron density at a fixed gate voltage depends on the magnetic field. 
We verify that this relationship holds over the entire gate voltage range. To do this, we plot in Fig.~\ref{Fig3}~(b) the dependence $n_{\rm QHE}(V_g)$ obtained in the same way as in Fig.~\ref{Fig2}~(c), but for magnetic fields ranging from 6 to 12\,T. It can be seen that all traces show linear dependences with the same slope, but shifted relative to each other. The vertical distance between the curves reflects the "excessive" electron density 
%$\Delta n = n_{\rm QHE} - (n-p)_{\rm SdH}^{\rm 2D}$ 
%$\Delta n = n_{\rm QHE} - n_{\rm SdH}^{\rm 2D}$ 
$\Delta n$
and is indeed linearly dependent on the magnetic field with a coefficient close to $e/h$ (Fig.~\ref{Fig3}~(c)). Extrapolating this trend to zero magnetic field, the dependence of $n_{\rm QHE}(V_g)|_{B = 0}$ agrees well with $n_{\rm SdH}^{\rm 2D}(V_g)$ shown in Fig.~\ref{Fig2}~(b) and (c). This  encourages us to define $\Delta n \equiv n_{\rm QHE} - n_{\rm SdH}^{\rm 2D}$, i.e. the difference between the charge density of 2D carriers in a given magnetic field and at $B = 0$.

Thus, the introduction of a magnetic field leads to the appearance of an excessive electron density $\Delta n$ close to $e|B|/h$. While maintaining the total density in the system, this can only be caused by the transition of carriers from the 3D bulk to the 2D layer near the gate. In turn, the decrease in the bulk electron density should enhance their  localization in magnetic field or even be the main cause of this process at $B>6$\,T. 
The exchange between quantized 2D carriers and a charge reservoir has been previously observed in HgTe QWs \cite{Kozlov2014b, Yakunin2020, Kozlov2023DF} as well as in graphene \cite{Kudrynskyi2017}, leading to the anomalous length of the Hall plateaus and known as a giant QHE.  
Although the manifestation of this phenomenon in our case is different, its reason remains the same and is associated with the different slope of the LLs formed by different carriers, which makes the transition energetically favorable. This assumption seems justified, since the presence of a single electron LL with a negative slope is a characteristic feature of HgTe QW with an inverted spectrum \cite{Scharf2012, Raichev2012}. To support our explanation numerically, we have performed the LLs calculations by using the eight-band $k\dot p$ Hamiltonian~\cite{Krishtopenko2016}. 
The calculations show (see 
%Sec.~\ref{SM_LL}
Sec.~S1
of SM~\cite{SM_universal}) that the slope of the first electron LLs associated with the surface states is smaller, then the slope of bulk LLs. A more detailed theoretical model of the exchange processes is currently lacking.

\begin{figure}
	\includegraphics[width=0.9\columnwidth]{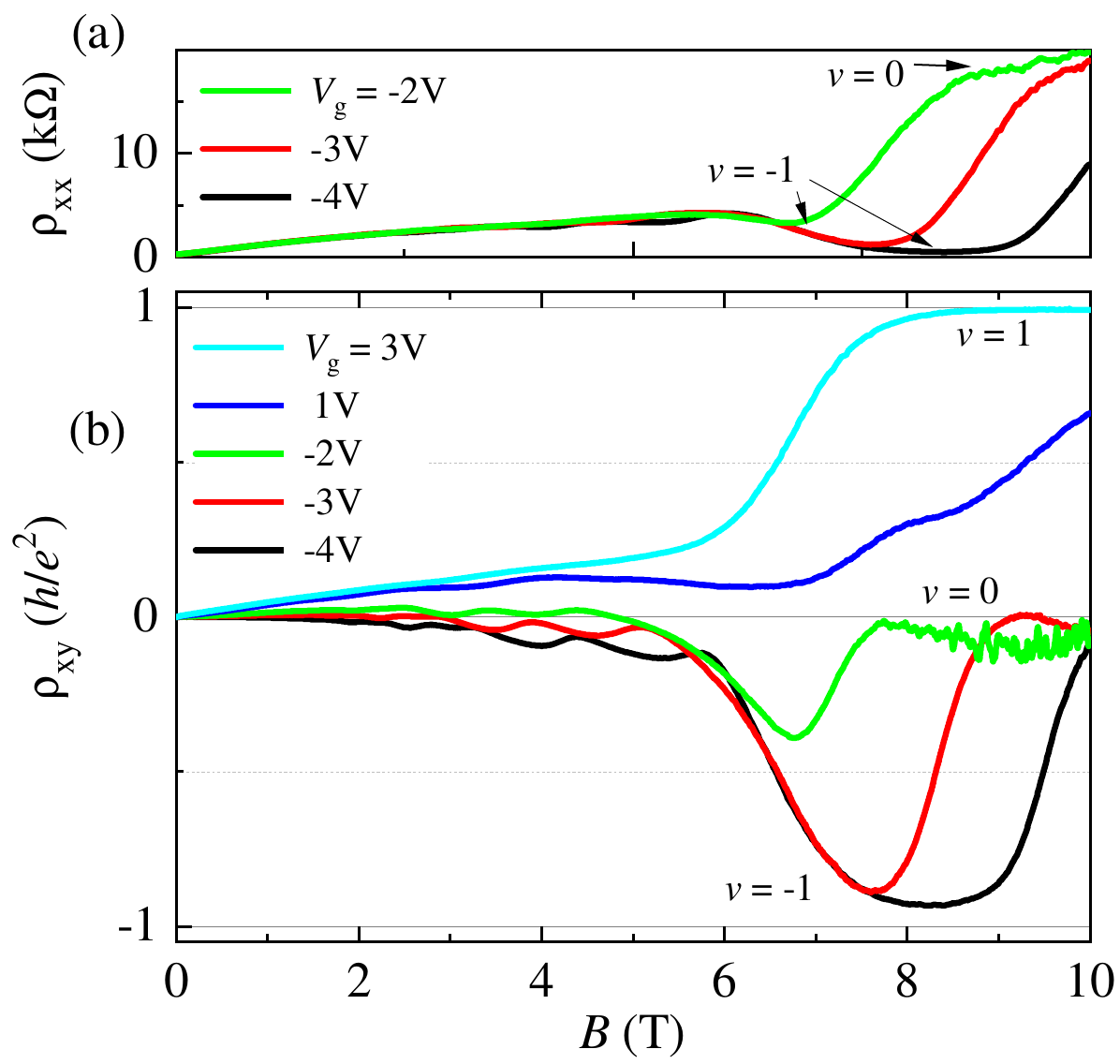}
	\caption{
		%Electron-hole scattering at $\nu=0$ and 
		%\textbf{Zero-plateau: its formation and the source of dissipation.}
		%Formation of zero-plateau
		%(a)~The illustration of electron-hole scattering in the regime of zero plateau: 
		%having higher electrochemical potential electrons 
		%1D electrons and holes move along the same edge and are equally biased by source-drain voltage that allows backscattering and strong dissipation of energy.
		(a) and (b)~The magnetic field dependences of $\rho_\text{xx}$ and $\rho_\text{xy}$ measured at different $V_\text{g}$.
		At lower fields, $\rho_\text{xy}$ goes to the first hole plateau $\nu=-1$, but as the field increases, the Hall resistance tends to zero values.
		%zero-plateau wins the game
		Note, the lower the voltage (and the lower the electron density), the higher the field is required to reach the zero plateau.
	} \label{Fig4}
\end{figure}

Finally, we analyze the process of zero plateau formation in a magnetic field. 
Fig.~\ref{Fig4} shows the dependence of $\rho_{\rm xx}(B)$ and $\rho_{\rm xy}(B)$ in the vicinity of the CNP, namely in the range $V_g = -4\ldots3$\,V. In weak magnetic fields, the value of $\rho_{\rm xy}$ turns out to be rather proportional to $(n-p)$, instead of $1/(n-p)$, due to the coexistence of electrons and holes contributing simultaneously to the conductivity~\cite{Savchenko2023}. In a strong magnetic field a transition to the QHE regime is expected. 
On the electron side, i.e. at $V_g = 3$\,V, a $\rho_{\rm xy}$ plateau corresponding to $\nu = 1$ is observed (Fig.~\ref{Fig4}~(b)) in the magnetic field of $9$\,T.
At first sight, the process of the hole plateau formation ($V_g = -4$\,V) is similar, but the almost completely formed hole plateau at $B = 8.5$\,T suddenly collapses at higher fields: the absolute value of $\rho_{\rm xy}$ decreases again and tends to zero.
The tendency becomes even more pronounced at $V_g = -3$. Finally, the zero plateau appears at $V_g = -2$\,V.
The same features are observed in the dependence $\rho_{\rm xx}(B)$, shown in Fig.~\ref{Fig4}~(a): first a minimum is formed corresponding to the filling factor $\nu = 1$, but then it is replaced by a maximum of about 20\,kOhm due to the dissipative nature of the edge transport at $\nu = 0$.
An attempt to form a QHE state with $\nu = 1$ and its transition to a stable  $\nu=0$ state is associated with the charge transfer between bulk and 2D carriers. Indeed, the gate voltage $V_g = -2$\,V corresponds to the center of the zero plateau at $B = 10$\,T and zero $n_{\rm QHE}$ density (Fig.~\ref{Fig2}~(c)), but in a zero magnetic field it also corresponds to the hole denisty $(p-n)_{\rm SdH}^{\rm 2D}\approx 2.5\cdot10^{11} $\,cm$^ {-2}$ (Fig.~\ref{Fig2}~(b)). Thus, as the magnetic field increases, conditions suitable for the formation of hole plateaus are replaced by conditions optimal for a zero plateau. At the same time, the concentration of bulk carriers decreases and their localization occurs.

\section{Outlook}
The presented observations reveal several directions and question for future theoretical and experimental studies regarding the QHE formation and possibilities to  utilize its unusially long-lasting chiral edge states.
First technical but interesting question, up to what thickness of a HgTe film is it possible to observe QHE?
We have found that here 3D carriers either tend to localize or move to the accumulation layer acquiring 2D nature as the magnetic field increases.
But there is no exact answer on if it is enough to realize QHE in much thicker HgTe films.
Moreover, it is still unclear which HgTe or the device property is crucial for the QHE formation -- its zero band gap, topological surface states with zero Landau level that has an anomalous magnetic field dependence or just the gate that facilitates the formation of a 2D near-gate conducting channel.
Additional high magnetic field studies in both HgTe crystals and other formally 3D systems should throw light on this general problem.

Another set of questions is related to the zero plateau in the Hall resistance.
Following our observations, we explain its formation by using an established theory of 1D QHE edge states and Landauer-Büttiker formalism.
The existing of such quasi-ballistic counter-propagating electron-hole edge channels on the sub-mm scale make this system a promising candidate to study their interaction and even possible exciton formation.

\section*{Acknowledgements}

N.N.M. and Z.D.K. acknowledge the support of RSF grant no. 23-72-30003. D.A.K. acknowledges the support of Dieter Weiss.

% Create the reference section using BibTeX:
\bibliography{library}

%\end{document}

%\documentclass[reprint, amsmath, amssymb, aps, superscriptaddress, prb, longbibliography, onecolumn]{revtex4-2}

%\usepackage{graphicx}% Include figure files
%\usepackage{epstopdf}
%\usepackage{dcolumn}% Align table columns on decimal point
%\usepackage{bm}% bold math
%\usepackage{color}
%\usepackage[colorlinks=true, urlcolor=blue, linkcolor=blue, citecolor=blue]{hyperref}
%\usepackage{natbib}
%\usepackage{amsbsy}
%\usepackage{lipsum}
%\usepackage{verbatim} 
%\usepackage[slovene]{babel}

%\usepackage{subfigure}
%\usepackage[dvipsnames]{xcolor}
%\usepackage{pst-all}
%\usepackage{ulem}\normalem

%\usepackage{lineno}
%\linenumbers

%\setcitestyle{super}
%\graphicspath{{figures/}{../figures/}}

%\begin{document}
\newpage
%\onecolumn
\onecolumngrid
\newpage

\section{Supplemental material to\\
Quantum Hall effect and zero plateau in bulk HgTe}

%\author{M.\,L.\,Savchenko}%\email{mlsavchenko@isp.nsc.ru}
%\affiliation{Institute of Solid State Physics, Vienna University of
	%Technology, 1040 Vienna, Austria}
%
 %\author{D.\,A.\,Kozlov}
%\affiliation{Experimental and Applied Physics, University of Regensburg, D-93040 Regensburg, Germany}
%\affiliation{Institute of Semiconductor Physics, 630090 Novosibirsk, Russia}
%
%
%
%\author{S.\,S.\,Krishtopenko}
%\affiliation{Laboratoire Charles Coulomb (L2C), UMR 5221 CNRS-Université de Montpellier, Montpellier, France}
%
%
%\author{N.\,N.\,Mikhailov}
%\affiliation{Institute of Semiconductor Physics, 630090 Novosibirsk, Russia}
%%\affiliation{Novosibirsk State University, Novosibirsk 630090, Russia}
%
%\author{S.\,A.\,Dvoretsky}
%\affiliation{Institute of Semiconductor Physics, 630090 Novosibirsk, Russia}
%
%\author{Z.\,D.\,Kvon}
%\affiliation{Institute of Semiconductor Physics, 630090 Novosibirsk, Russia}
%\affiliation{Novosibirsk State University, Novosibirsk 630090, Russia}
%
%\author{A.\,Pimenov}
%\affiliation{Institute of Solid State Physics, Vienna University of
	%Technology, 1040 Vienna, Austria}
%
 %\author{D.\,Weiss}
%\affiliation{Experimental and Applied Physics, University of Regensburg, D-93040 Regensburg, Germany}
%
%\date{\today}
%
%
%\maketitle
%
\setcounter{figure}{0}
\setcounter{equation}{0}
\renewcommand{\thesection}{S\arabic{section}}
\renewcommand{\theequation} {S\arabic{equation}}
\renewcommand{\thefigure} {S\arabic{figure}}
\renewcommand{\thetable} {S\arabic{table}}

\subsection{Landau level calculations}

\begin{figure}[h]
	\includegraphics[width=0.5\columnwidth]{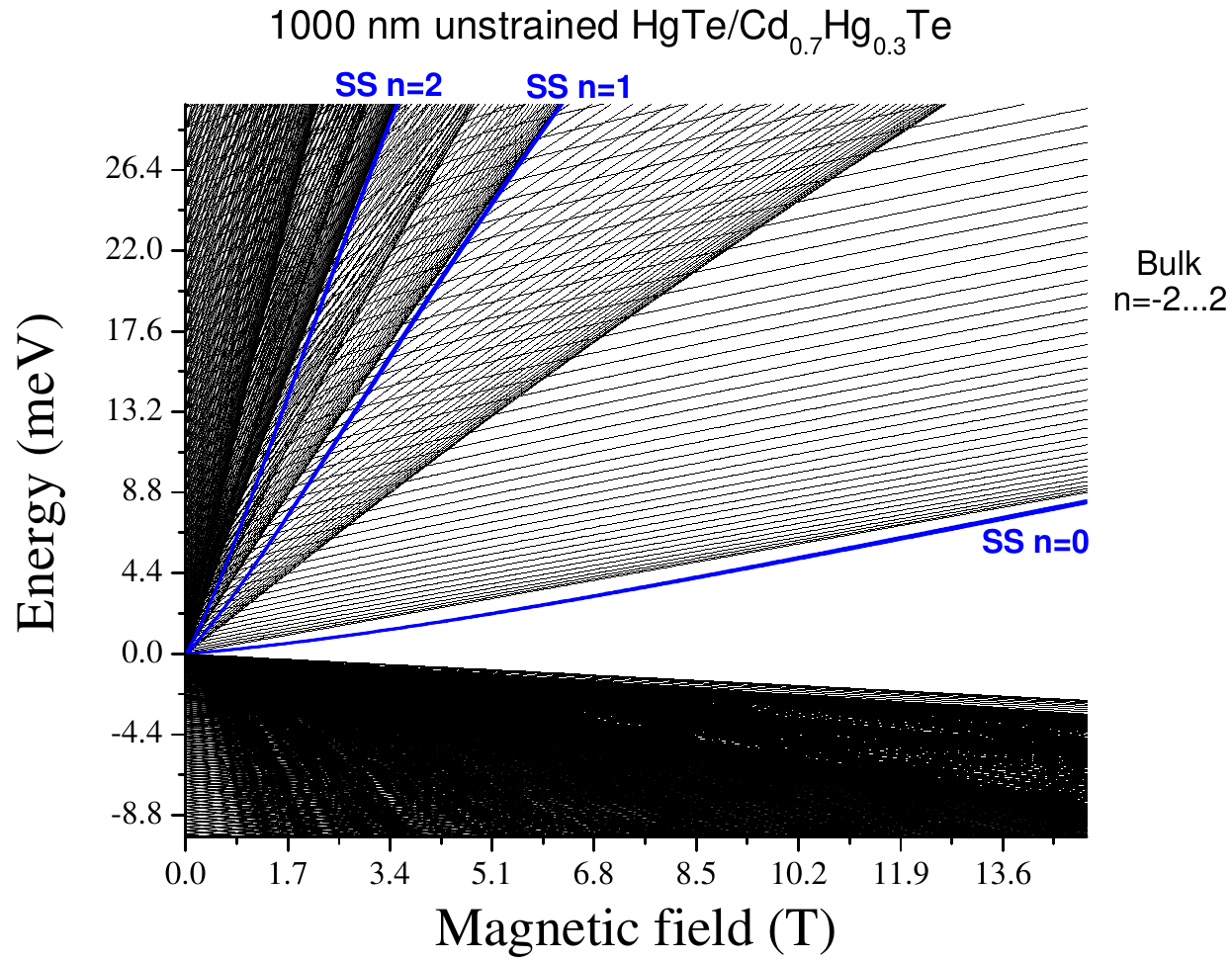}
	\caption{
		Energy of Landau levels as a function of magnetic field in a 1000 nm (013) HgTe film surrounded by Cd$_{0.7}$Hg$_{0.3}$Te barriers. The magnetic field is oriented along the growth direction. 
        Black curves correspond to the Landau levels of the bulk states with the level indices $n=-2$, $-1, \dots, 2$. 
        Blue curves represent the  Landau levels of surface states (SS) localized in the vicinity of the HgTe/Cd$_{0.7}$Hg$_{0.3}$Te heterojunctions. 
        For the notations of Landau levels, see~\cite{Krishtopenko2016}.
	} \label{SM_LL}
\end{figure}

To support our interpretation numerically, we have performed the LLs calculations by using the eight-band $k\cdot p$ Hamiltonian~\cite{Krishtopenko2016}.  The latter directly takes into account the interactions between $\Gamma_6$, $\Gamma_8$ and $\Gamma_7$ bands in bulk materials, resulting in a good agreement between theoretical calculations and experimental results in HgTe quantum wells~\cite{Ikonnikov2016, Marcinkiewicz2017, Mikhailov2018}. 
The calculations have been performed by expanding the eight-component envelope wave functions in the basis set of plane waves and by numerical solution of the eigenvalue problem. Details of calculations, the form of the Hamiltonian can be found elsewhere~\cite{Krishtopenko2016}. Parameters for the bulk materials, and valence band offsets used in the calculations are also provided in~\cite{Krishtopenko2016}. 
Fig.~\ref{SM_LL} shows the calculation results for the sample under study as a function of $B$. One could clearly see, that the slope of the first electron LLs associated with the surface states is smaller, then the slope of bulk LLs. The calculation does not take into account the presence of accumulation layer. A more detailed theoretical model of these processes is currently lacking. In particular, it is unknown what happens to the "excessive" electrons at negative gate voltage: both coexistence with 2D holes and their mutual compensation are possible.

\newpage
\subsection{QHE in local and non-local resistance, and in conductivity}
\label{SM_QHE_and_Non-local}
\noindent

\begin{figure*}[h]
	\includegraphics[width=0.7\columnwidth]{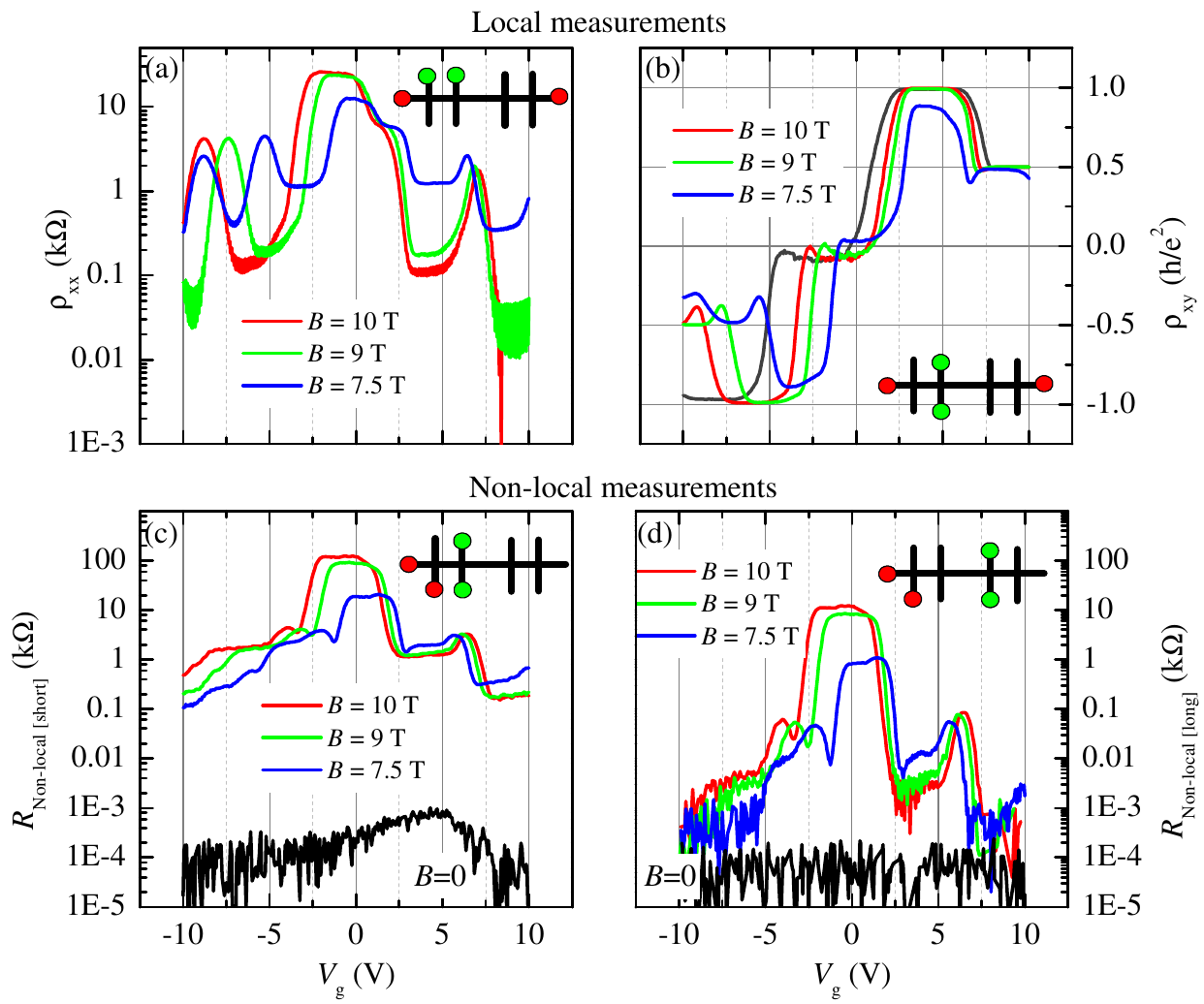}
	\caption{
		(a) and (b)~QHE plateaus in Hall resistance and minima in longitudinal resistance  measured at different magnetic fields and at temperature $T = 0.16\,$K. 
        (c) and (d)~Non-local resistance versus gate voltage measured at  same magnetic fields and temperature in different configurations. 
        Insets: the measurement scheme, red and green dots represent the current and voltage probes, respectively.
	} \label{fig_SM_QHE_and_Non-local2_v2}
\end{figure*}

\begin{figure*}[h]
	\includegraphics[width=0.65\columnwidth]{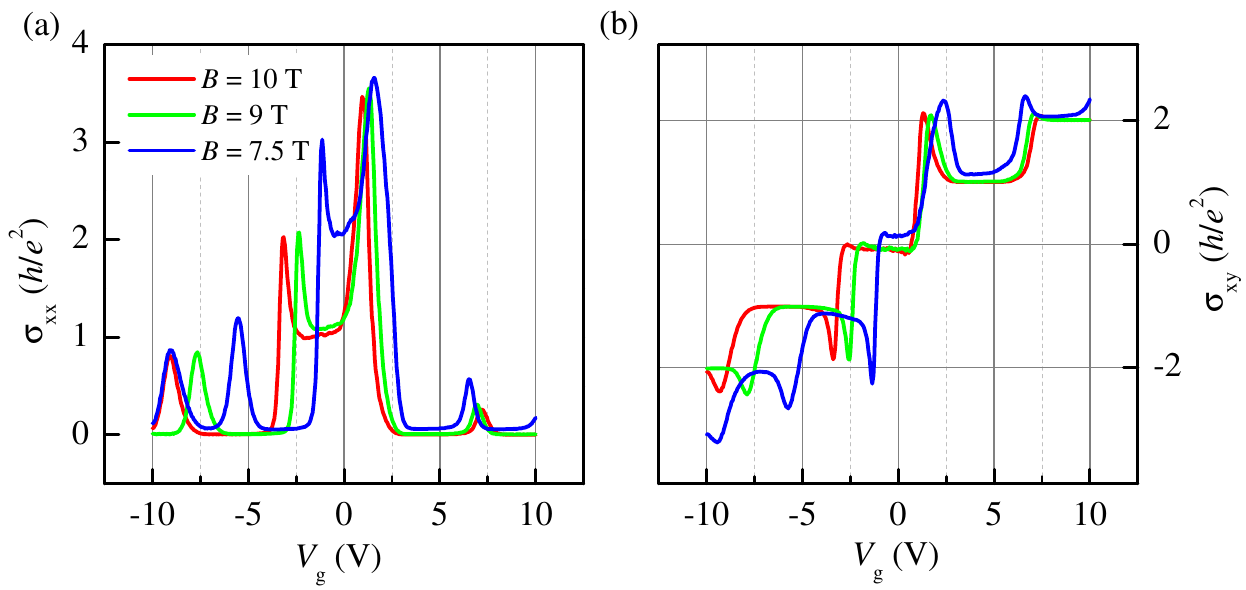}
	\caption{
		Recalculated in conductivity curves from Fig.~\ref{fig_SM_QHE_and_Non-local2_v2}~(a) and (b).
	} \label{fig_SM_QHE_sigma}
\end{figure*}

\newpage
\subsection{Zero plateau before and after data processing}
\label{sec: SM_Rxy asymmetry}
\noindent

\begin{figure*}[h]
	\includegraphics[width=0.6\columnwidth]{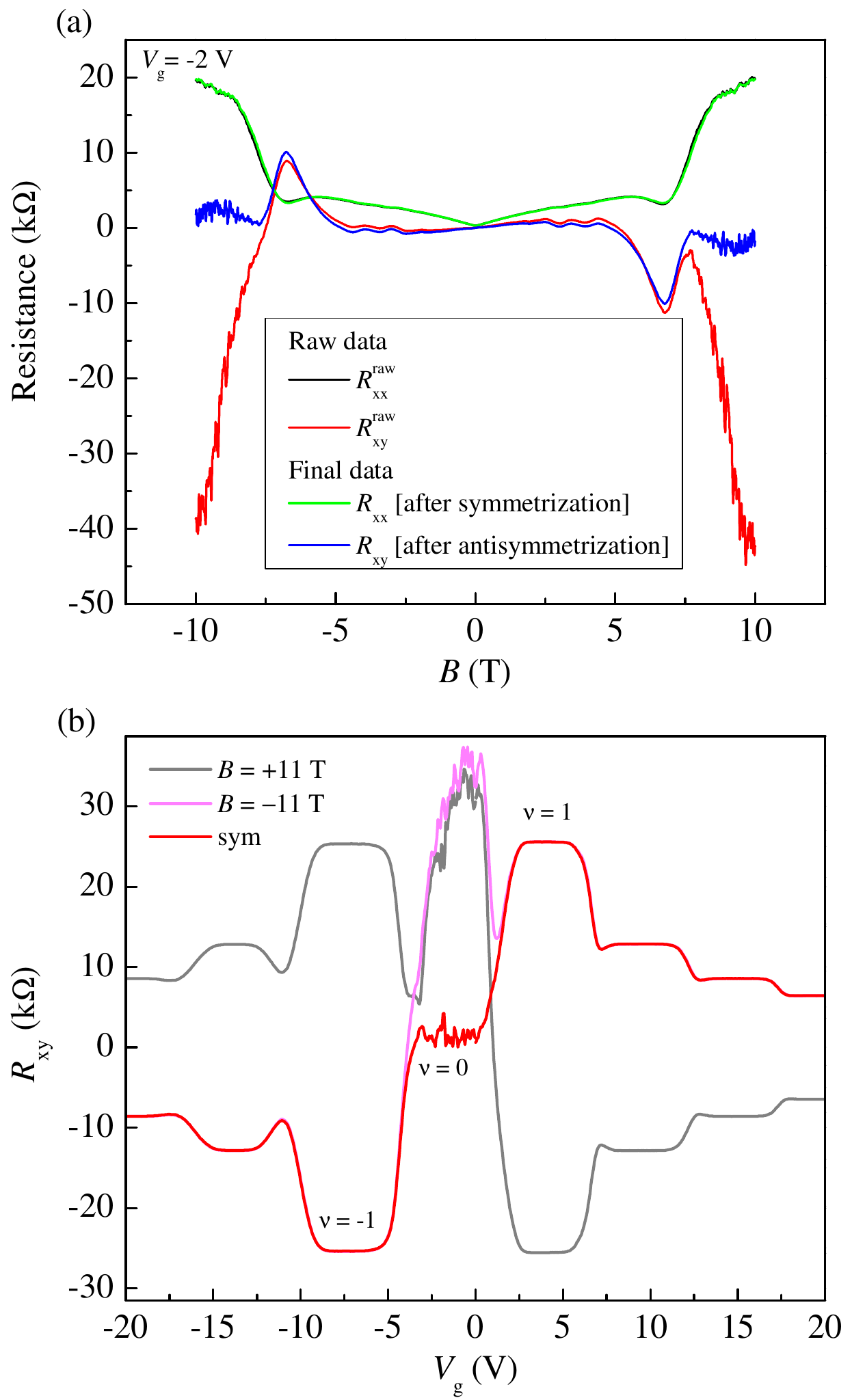}
	\caption{
		(a)~The magnetic field dependences of the diagonal and Hall resistance measured at $V_\text{g} = -2$~V before and after data processing,
        $R_\text{xx}(B) = \frac{R_\text{xx}^\text{raw}(B) + R_\text{xx}^\text{raw}(-B)}{2}$,
        $R_\text{xy}(B) = \frac{R_\text{xy}^\text{raw}(B) - R_\text{xy}^\text{raw}(-B)}{2}$.
%(a)~The magnetic field dependences of the diagonal ans Hall resistance $R_\text{xx}$ and $R_\text{xy}$ measured (black and red) at $V_\text{g} = -2$~V.   
        (b)~The gate voltage dependences of the Hall resistance in the QHE regime before (grey and pink) and after (red) the antisymmetrization procedure.
	} \label{SM_Rxy_Rxx_vs_B_m2V}
\end{figure*}

\newpage
\subsection{QHE at different temperatures and fields}
\label{sec: QHE_diff_T}
\noindent
\begin{figure*}[h]
	\includegraphics[width=0.8\columnwidth]{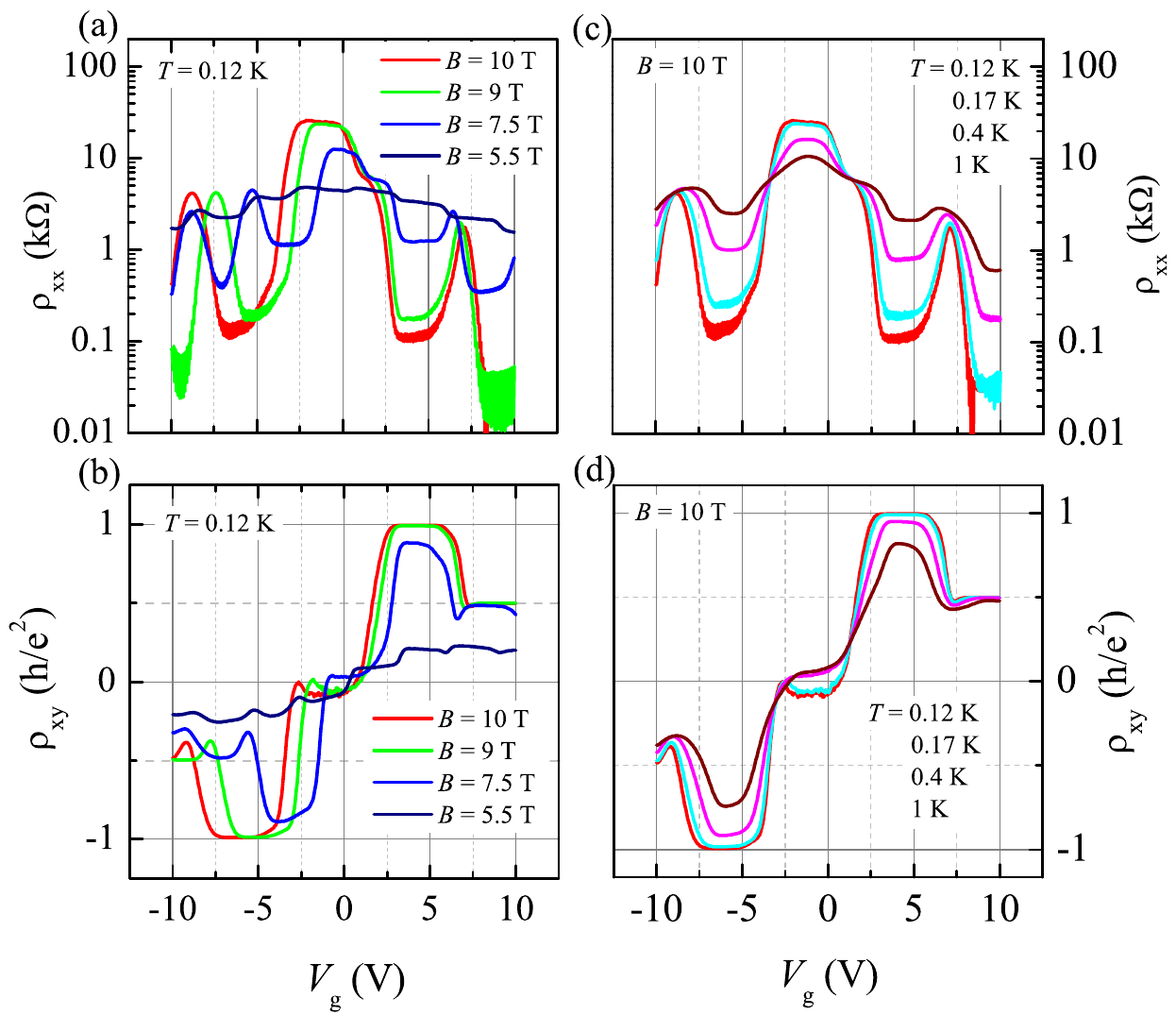}
	\caption{
		The gate voltage dependences of $\rho_\text{xx}$ (top panels) and $\rho_\text{xy}$ (bottom panels).
		Left panels (a) and (b) correspond to different magnetic fields and $T = 0.12\,$K.
		Right panels (c) and (d) correspond to different temperatures and $B = 10\,$T.
	} \label{Fig_QHE_diff_T}
\end{figure*}
%

%\bibliography{library} 

\end{document}